\documentclass{llncs}
\usepackage[pdftex]{graphics}
\usepackage{multicol}
\usepackage{latexsym}
\usepackage{amsmath}
\usepackage{amstext} 
\newsavebox{\savepar}

\usepackage{amssymb} 
\usepackage{theorem}
\newcommand{\function}[3]{#1 : #2 \mapsto #3}
\newtheorem{Thm}{Theorem}

\newtheorem{Defi}[Thm]{Definition}
\newtheorem{Exa}[Thm]{Example}

\newtheorem{Rem}[Thm]{Remark}
{\end{trivlist}}

\begin{document}
\title{A study of fuzzy and many-valued logics in cellular automata}\vskip0.1cm
\author{Angelo B. Mingarelli}
\institute{School of Mathematics and Statistics, Carleton
University, Ottawa, Ontario, Canada, K1S 5B6:  
amingare@math.carleton.ca.}
\maketitle
\section{Introduction} 

In this paper we provide an analytical study of the theory of multi-valued and fuzzy cellular automata where the fuzziness appears as the result of the application of an underlying multi-valued or continuous logic as opposed to standard logic as used conventionally, e.g., \cite{nks}.  Thus, we consider boolean elementary cellular automata (ECA) and in particular those CA considered by Wolfram \cite{nks} as a starting point. For an excellent review of classical CA theory and current problems within see \cite{mm}. Using the disjunctive normal form of any one of the 255 ECA's so defined, or {\it rules} as we call them sometimes,  we modify the underlying logic structure and redefine the ECA within the framework of this new logic. This then defines a CA under a new logic: For example, the classical ECA \cite{nks}, are CA under boolean (two-valued) logic, the fuzzy cellular automata (FCA) defined in \cite{cat1} and considered in \cite{cat2},  \cite{rule90}, \cite{FloSan94a}, \cite{amca1}, \cite{amca2}, \cite{abmf}, \cite{ijuc}, are continuous CA under a special fuzzy logic (see Sections 3-4 herein). This latter logic has been used recently to generalize the game of {\it Life}, see e.g., \cite{reiter}, and is used to investigate the result of perturbations, noisy sources, computation errors, mutations, on the evolution of boolean ECA, \cite{FloSan94a}.  In addition, since all electromagnetic signals travel at finite speeds, there is a point at which a boolean switch is neither ON nor OFF. Although this time interval is very short, this can be considered a fuzzy event, that is, an event where the switch is neither on nor off and is therefore in a countable number of intermediate states. As miniaturization and nanoscale technology becomes the norm, we suspect that such fuzzy or continuous states will play an important role in the development of future technologies, where quantum mechanical laws will intermingle with fuzzy states continuously and may even coincide.
\vskip0.15in

The idea here is to study the role of the logical structure on the dynamics of the new automaton and to determine whether or not complexity exists; in particular, the main question asked is: `` Are the dynamics of the resulting space-time diagrams chaotic?" This is, of course, a difficult question to answer in this generality so we will  target an important CA namely, ECA 110, and study its space-time diagram under a change of logic from boolean to probabilistic.  Although it is somewhat surprising that FCA 110 is incapable of chaotic dynamics (see \cite{amca1})  as opposed to its boolean counterpart \cite{nks}, it is definitely odd that the present study of a special probabilistic CA known as PCA 110 also appears to lean in this direction. One would expect that of all the continuous logics available, an underlying {\it probabilistic} logic (see Example 2) should yield chaotic space-time diagrams for the corresponding cellular automaton to boolean rule 110, however, such a premise is not supported by theoretical arguments. The ultimate point of all this is to determine whether such non-chaotic evolution can shed some light into the sometimes chaotic evolution of the same automaton, under boolean logic,  as the seed value approaches the number 1 from the left. We suspect that a necessary condition for the existence of chaotic space-time diagrams for a general fuzzy rule is that the underlying logic be discontinuous as a function of the two variables, in the sense of mathematical analysis. 
\vskip0.15in 

We will see that if the seed $a$, in a homogeneous background of zeros, is chosen sufficiently close to but less than $1$, then the {\it space-time diagrams}  (that is the space vs. time evolution of the automaton)  of both boolean and fuzzy rule 110 (denoted by FCA 110) up to time $T$ will be {\it identical}. Thus, one may expect that the asymptotic evolution of the fuzzy space-time diagram will have a bearing on the distribution of $0$'s and $1$'s in the corresponding asymptotic boolean evolution. More precisely, we can say that given any row number, or time,  $T>0$, there exists a seed value $a < 1$ such that the  evolution of ECA 110 under a specific fuzzy logic (dubbed CFMS logic, see below) up to row $T$ is identical to its boolean evolution up to row $T$ (provided we color all cells having values greater than one-half as black and all others as white) while for all subsequent rows the evolution is {\it non-chaotic} whereby various limits exist and are given by quotients of linear combinations of Fibonacci numbers, see \cite{amca1} for specific details and the sequel for examples. We know that under CFMS logic all but 6 of the 255 ECA admit non-chaotic behavior for any finite random initial string \cite{abmf}, \cite{ijuc}. A similar statement can be formulated for probabilistic ECA 110 (denoted by PCA 110) but the precise result is conjectural at this time due to the massive amount of technical details that arise in its investigation (see the last section). We suspect that those CA that lead to truly random behavior (whether classical ECA or fuzzy) are those whose space-time evolution from one seed gives limits that are one-half, independently of the seed value. For example, ECA 18 and its fuzzy counterpart, FCA 18 (see below) would enter into this realm. We can only surmise, and cannot prove completely, that PCA 110 is non-chaotic as well on finite initial strings. We leave it as an open question the problem of determining whether there is non-chaoticity (lack of complexity) in ECA under the other multi-valued logics mentioned here (Lukasiewicz, etc.)

\section{Logical structures and cellular automata}

\vskip0.15in 

Basically, a logic is {\it multi-valued} if it takes more than two values (e.g., more than just True or False, 1 or 0,  etc.). Thus, by definition, multi-valued logics allow for truth-values that are more than just {\it true} or {\it false}. In a sense they allow values {\it in between}, which is where the fuzziness concept comes in. For example, in classical propositional logic the law of the {\it excluded middle}  states that for any proposition P,  the statement ($P \vee \neg P$) (P or not P) is always true. Such a statement allows for ``proofs by contradiction" which is a technique that cannot be used generally  for multi-valued logics. Since its origins in the 1920's there has been considerable work in multi-valued logics, in the past 20 years much of it resulting from the application of fuzzy techniques to real-world problems (cf., e.g., \cite{rt}, \cite{klir2}). For generalities concerning {\it fuzzy} automata we refer the reader to [\cite{fss}, p.199] while we cite \cite{esko} for details concerning fuzzy logics. Our aim here though is to entertain the use of general many-valued or continuous logics as well as general CA  and to single out one or two of them in order to show that some fuzzy logics can give rise to non-chaotic behavior using the corresponding ECA. For various applications of many-valued logics to hardware design and artificial intelligence, see \cite{got}, \cite{glu}, \cite{maji}.

Whenever a logic takes on exactly three values it is called {\it trivalent} [\cite{klir}, p.192]. The three values may take the form True, False, and something in between or we can also use numbers such as 1, 0, and 1/2 (cf., \cite{rt}). A logic that takes on finitely many values is called an {\it n-valued logic} while an infinitely-valued logic is one that takes on infinitely many values. Specifically, a {\it continuous logic} is one that takes on a continuum of values. In particular, in this article we will always assume that our logics  have their values in the closed interval $[0,1]$ of real numbers. 
\vskip0.15in
Denoting the two-point set (or boolean space) consisting of the numbers $0$ and $1$ by $\{0,1\}$, classical propositional (or standard) logic gives the truth tables

\begin{table}[h]
\begin{center}
\begin{tabular}{|c|c|c|}\hline
$\mathbf{P \vee Q} $& $P=0$ & $P=1$ \\ \hline
$Q=0$ & $0$ & $1$  \\ \hline
$Q=1$ & $1$ & $1$  \\ \hline
\end{tabular}\,\,\,\,\,\,\,\,\,\,\,
\begin{tabular}{|c|c|c|}\hline
$\mathbf{P \wedge Q} $& $P=0$ & $P=1$ \\ \hline
$Q=0$ & $0$ & $0$  \\ \hline
$Q=1$ & $0$ & $1$  \\ \hline
\end{tabular}\,\,\,\,\,\,\,\,\,\,\,
\begin{tabular}{|c|c|c|}\hline
$\mathbf{P \Rightarrow Q} $& $P=0$ & $P=1$ \\ \hline
$Q=0$ & $1$ & $0$  \\ \hline
$Q=1$ & $1$ & $1$  \\ \hline
\end{tabular}
\end{center}
\end{table}

\noindent{t}he truth table of $\neg P$ being found easily because of the law of the excluded middle.  Recall that $P \Rightarrow Q$  is defined classically as $(\neg P) \vee Q$. Prior to undertaking any constructions we illustrate by means of examples some of the multi-valued/continuous logics we are considering in this work.
\begin{Exa} Some common multi-valued logics and their formulations can be seen in the following two tables: The Lukasiewicz p-logic is defined for all $p \geq 1$, the case $p=1$ being the most commonly used one. Since these logics have their values in the closed interval $[0,1]$, $x,y \in [0,1]$ therein.

\begin{table}[h]
\begin{center}
\begin{tabular}{|c||c|c|c|}\hline
$\mathbf{} $& Lukasiewicz\, p-logic & Zadeh logic  & CFMS logic\\ \hline 
$\mathbf{\neg x} $ & $1-x$ & $1-x$  & $1-x$\\ \hline
$\mathbf{x \wedge y} $ & $\sqrt[p]{\max \{0, x^p + y^p -1\}}$ & $\min\{x,y\}$ & $x\cdot y$ \\ \hline  
$\mathbf{x \vee y} $ & $\min\{1, \sqrt[p]{x^p + y^p}\}$ & $\max\{x,y\}$ & $\min\{1, x+y\}$ \\ \hline 
$\mathbf{x \Rightarrow y} $ & if $x\leq y$ then 1, else $1- x^p + y^p$ & $\max\{1-x,y\}$ & $\min\{1, 1-x+y\}$ \\ \hline 
\end{tabular}
\end{center}
\end{table}
\end{Exa}


\begin{table}
\begin{center}
\begin{tabular}{|c||c|c|c|}\hline
$\mathbf{} $&G\"{o}del logic & Product logic  & Probabilistic logic\\ \hline 
$\mathbf{\neg x} $ & if $x=0$ then 1 else 0 &  if $x=0$ then 1 else 0  & $1-x$\\ \hline
$\mathbf{x \wedge y} $ & $\min\{x,y\}$ & $x\cdot y$ & $x\cdot y$ \\ \hline  
$\mathbf{x \vee y} $ & $\max\{x,y\}$ & $x + y - x\cdot y$ & $x + y - x\cdot y$ \\ \hline 
$\mathbf{x \Rightarrow y} $ & if $x\leq y$ then 1, else $y$ & if $x\leq y$ then 1, else $x/y$& $y - x + x\cdot y$ \\ \hline 
\end{tabular}
\end{center}
\end{table}

\begin{remark} \noindent{T}he logic described in the table below as CFMS logic seems to appear first in the paper \cite{cat1}. Observe that the functions defined by $\neg x$ on $[0,1]$ and $x \wedge y$, $x \vee y$ on $[0,1]^2$ map back into $[0,1]$ for each logic presented. However, each of these functions is continuous on $[0,1]$ or $[0,1]^2$ only when the underlying logic is neither a G\"{o}del or Product logic. The {\it implication} function on $[0,1]^2$ is continuous only for the Zadeh, CFMS or Probabilistic logic. Thus, although each of these 6 logics is a continuous logic, only a few of them define continuous functions on $[0,1]^2$, {\it per se}. This last statement has a bearing on the complexity of cellular automata based on multi-valued logics.
\end{remark}

We begin by defining simple boolean (two state) automata based on standard logic. A (boolean)  {\it cellular automaton} is a collection of black and white cells (or ones (1) and zeros (0) resp., or true (T) or false (F) statements, resp.) arranged on an infinite one-dimensional strip. We also call this an {\it initial string}. Each cell is preceded by a cell behind it and a cell in front of it; these are called the {\it neighbors} of the cell.  We distinguish an arbitrary (but fixed) special element of this strip by labeling it $x_0$ (or by $x_{0}^{0}$). Returning to our band of initial values (or our initial configuration) we define a {\it local rule}, by a function $\function{g}{\{0,1\}^3}{\{0,1\}}$, on consecutive triples $x_{i-1}, x_i, x_{i+1}$ of this band with the effect that  $f(x)_i = g(x_{i-1}, x_i, x_{i+1})$ where the value $f(x)_i$ is placed directly below $x_i$ (on another such parallel band) and this update of the original cell values is performed for all the quantities in the initial string resulting in two parallel bands. The procedure is then repeated to generate an infinite sequence of bands whose structure at infinity we wish to study. Ultimately, the idea is to compare and contrast the effect of the various logics used in determining the long term evolution of cellular automata based upon them; thus the motivation for calling these continuous or fuzzy cellular automata.

We now assume standard propositional calculus. Thus, when dealing with (classical) boolean CA's (as in \cite{nks})  the {\em local rule} is defined by the 8 possible
local configurations of triples that a cell detects in its neighborhood:
$$000, 001, 010, 011, 100, 101, 110, 111 \rightarrow
r_0, r_1, \cdots,
r_7$$
where each triplet represents a local configuration of
the left neighbor,
the cell itself, and the right neighbor, and the numbers $r_i$ are either $0$ or $1$. Letting the $r_i$ take on all possible values of $0$'s and $1$'s will result in the definition of $2^8 = 256$ different local functions. Each one of these local functions defines a rule whose rule name is, by definition, the value of the sum 
 $${\rm Rule\ \ name}\, \equiv \sum_{i=0}^{7} r_i\, 2^i, $$ see Examples~\ref{ex184}, \ref{ru18}, and \ref{ex30}  below. Now it is easy to see that our definition of rule names forces a given rule name to be an integer between 0 and 255 and, indeed, every such integer defines a rule name. Observe that, by definition, the string $r_7r_6r_5\ldots r_1r_0$ is simply the binary representation of the rule name (that is, the representation of that integer in base 2).

Now, it is known that every local rule of a boolean CA is
expressible as a {\em  disjunctive normal form} (abbr. DNF), i.e.,
$$g(x_1,x_2, x_3) = \vee_{i|r_i=1} \wedge_{j=1}^{3}
x_j^{d_{ij}}$$
where $d_{ij}$ is the $j$-th digit, from left to
right, of the binary
representation of $i$, and $x^0$ (resp. $x^1$) stands for
$\neg x$ (resp.
$x$). This DNF is simply a comprehensive device for the writing of any one of the 255 general boolean CA's using the standard conjunctions and disjunctions of classical propositional logic. More precisely, the DNF is the form to which every statement of propositional calculus can be reduced, consisting of a disjunction of conjunctions each of the conjuncts of which is either an atomic formula or its negation.

This representation of $g$ is also called the {\it canonical expression} for the rule. For example, since $3=1\cdot 2^0 + 1\cdot 2^1+0 \cdot 2^2 +0 \ldots$, we can write $3=(011)$ in base 2, so that $d_{31}=0, d_{32}=1, d_{33}=1.$ Similarly, since $5=1\cdot 2^0 + 0\cdot 2^1+1 \cdot 2^2 +0 \ldots$, we have $5=(101)$ in base 2, and $d_{51}=1, d_{52}=0, d_{53}=1.$ In the same way we can show that, for example, $d_{41}=1, d_{42}=0, d_{43}=0$ and $d_{71}=1, d_{72}=1, d_{73}=1$  (these quantities are used in Example~\ref{ex184} below).

In addition, the rule $g$ which sends the boolean triples $$000, 001, 010, 011, 100, 101, 110, 111 \rightarrow
0,1,1,0,0,1,0,1$$
is a local rule. Since the effect of this rule on every such triple may be written as $g(1,0,1)=1$, $g(0,1,0)=1$, $g(1,0,0)=0$, $g(0,0,0)= 0$, etc., this rule maps the initial string, say,
\begin{eqnarray*}
\ldots 1\quad 0\quad 1 \quad 0 \quad 0 \quad 1\quad 1 \quad 1 \ldots
\end{eqnarray*}
into (the second band in the next display, and then the third band, etc.)
\begin{eqnarray*}
&&\ldots 1\quad 0\quad 1 \quad 0 \quad 0 \quad 1\quad 1 \quad 1 \ldots\\
 &&\ldots\ldots\,1\quad 1 \quad 0 \quad 1 \quad 0\quad 1 \quad \ldots \\
&&\ldots\ldots\ldots\,\,0 \quad 1\quad 1 \quad 1\quad \ldots {\rm \&\, etc}.
\end{eqnarray*}
This rule $g$ is an example of a boolean CA and it is such boolean CA's (see \cite{nks}) and special multi-valued/fuzzy counterparts to be defined below that interests us here as an application of the general theory that is developed in \cite{abmf}, \cite{ijuc} in the case of an underlying natural fuzzy logic. The {\it global dynamics} (or evolution or asymptotics) of a rule is defined by updating the initial and all subsequent string values according to a repeated application of the local function to the neighborhood of each cell. 

\begin{Exa}
\label{r110}
Since $110 = 2+2^2+2^3+2^5+2^6$ we see that boolean rule 110 where $110 = \sum_{i=0}^{7} r_i\,2^i $ forces $r_i = 1$ only for $i=1,2,3,5,6$. Use of the disjunctive normal form expression above gives us
\begin{eqnarray}
g_{110}(x_1,x_2, x_3) & = &\vee_{i|r_i=1} \wedge_{j=1}^{3} x_j^{d_{ij}},\nonumber\\
&=&  (x_1^{d_{11}} \wedge x_2^{d_{12}} \wedge x_3^{d_{13}}) \vee (x_1^{d_{21}} \wedge x_2^{d_{22}} \wedge x_3^{d_{23}}) \vee  (x_1^{d_{31}} \wedge x_2^{d_{32}} \wedge x_3^{d_{33}})\nonumber \\
&& \vee (x_1^{d_{51}} \wedge x_2^{d_{52}} \wedge x_j^{d_{53}}) \vee  (x_1^{d_{61}} \wedge x_2^{d_{62}} \wedge x_3^{d_{63}})\nonumber \\
& = & (x_1^{{0}} \wedge x_2^{{0}} \wedge x_3^{{1}}) \vee (x_1^{{0}} \wedge x_2^{{1}} \wedge x_3^{{0}}) \vee (x_1^{{0}} \wedge x_2^{{1}} \wedge x_3^{{1}}) \vee (x_1^{{1}} \wedge x_2^{{0}} \wedge x_3^{{1}}), \nonumber \\
&& \vee (x_1^{1} \wedge x_2^{1} \wedge x_3^{0}) \nonumber \\
& = & (\neg x_1 \wedge  \neg x_2 \wedge x_3) \vee (  \neg x_1 \wedge  x_2 \wedge  \neg x_3) 
\vee ( \neg x_1 \wedge x_2 \wedge  x_3) \vee \nonumber \\
 && \vee (  x_1 \wedge \neg x_2 \wedge  x_3) \vee (  x_1 \wedge x_2 \wedge  \neg x_3) \label{b1}
\end{eqnarray}
\end{Exa}
for $x_i \in \{0,1\}$, $i=1,2,3,$ in the case of a boolean logic or $x_i \in [0,1]$, $i=1,2,3,$ in the case of a multi-valued/fuzzy logic. In the sequel we will assume that (\ref{b1}) is the canonical form of ECA 110 devoid of any underlying logical structure. 
\vskip0.15in
One can now ask the following natural question: {\it Under what conditions is ECA 110 under a finitely-valued logic universal?} The corresponding result for classical (two-valued logic) is believed by everyone (cf., [\cite{nks}, p.865]), but this is not clear for even a trivalent logic. The point is that a general method for determining the universality of such an ECA under any logic is lacking.

\section{ECA 110 under continuous or fuzzy logics}

We motivate {\it continuous} or {\it fuzzy} cellular automata (or continuous ECA defined by them) by means of an example. In this case we note that the boolean space consisting of the two points 0 and 1, that is, $\{0, 1\}$, is now replaced by an interval, namely, the closed interval of points between 0 and 1, denoted by $[0,1]$. 

\begin{Exa} (Fuzzy rule 110 also known as FCA 110) Consider the continuous rule $g$ defined by $g(x_1,x_2, x_3)=  x_2+x_3-x_2x_3-x_1x_2x_3$, for $x_i \in [0,1], i=1,2,3.$ This is, in fact what we call {\it fuzzy rule 110} in \cite{amca1}. This expression for $g$ is obtained by casting boolean ECA 110 (see Example~\ref{r110} above) under  CFMS logic (this process is called fuzzification). That is, in the DNF we redefine for real numbers $a,b \in [0,1]$, the quantities $(a \vee b)$   to be $(a+b)$, $(a \wedge b)$ to be $(ab)$, and $(\neg a)$ to be  $(1-a)$, in accordance with CFMS logic. In other words, $a \vee b = a+b$, $a \wedge b = a\cdot b$, and $\neg a = 1-a$, where $+$ and ``$\cdot$" are ordinary addition and multiplication of real numbers. Specifically, starting from the DNF of ECA 110, we find (after fuzzification) 
\begin{eqnarray}
g^{F}_{110}(x_1,x_2,x_3) & = & (\neg x_1 \wedge  \neg x_2 \wedge x_3) \vee (  \neg x_1 \wedge  x_2 \wedge  \neg x_3) 
\vee ( \neg x_1 \wedge x_2 \wedge  x_3) \vee \nonumber \\
&& \vee (  x_1 \wedge \neg x_2 \wedge  x_3) \vee (  x_1 \wedge x_2 \wedge  \neg x_3) \nonumber \\
& = & (1-x_1)(1-x_2)x_3 + (1-x_1)x_2(1-x_3)+(1-x_1)x_2x_3 +\nonumber \\
&&+ x_1(1-x_2)x_3 + x_1x_2(1-x_3), \nonumber \\
& = & x_2+x_3-x_2x_3-x_1x_2x_3.
\end{eqnarray}
\end{Exa}

As an example of the long-term dynamics of this rule, we let this FCA act on the finite string consisting of a singleton $a=0.426$ in a background of zeros (i.e., ``$a$" is surrounded on both sides by zeros, see Table~\ref{t4} below). Since $g(0,0,0)=0$ the value below the central ``0" of any string of the form 0,0,0 in Table~\ref{t4} is zero. Next, $g(0,0,.426)=.426$ so this value is placed under the central zero of the triple $0,0,.426$. Next, $g(.426, .670, .426) = 0.689$. Hence this value is placed in the entry under the cell value $0.670$. Similarly, $g(.937, .700, .510) = .518$ (which goes under $0.7$) and the process is repeated indefinitely for all possible cell values in every band. This generates an infinite number of bands consisting of the ``window" of values below in Table~\ref{t4} (also called the {\it space-time diagram} of the automaton). The basic question is, ``What happens to the cell values when we look at arbitrarily large number of such bands?" Do the cell values converge? If so, to what? Is there chaos?
\begin{table}[h]
\begin{center}
\begin{tabular}{ c c  c  c  c  c c c c c  c  c  c}	\hline
$\ldots 0$ & $0$ &$0$ &$0$ &$\mathbf{ 0}$&$\mathbf{0}$&$\mathbf{0.426}$ & $0$ & $0$ &$0$ &$0$ & 0&0\ldots \\	\hline
$\ldots 0$ & $0$ &$0$ &$0$ &$ 0$&$\mathbf{0.426}$&$ .426$ & $0$ & $0$ &$0$ &$0$ & 0&0\ldots \\	
$\mathbf{ 0}$ & $\mathbf{ 0}$ &$\mathbf{ 0}$ &$0$ & .426&$.670$&.426    & $0$ & $0$ &$0$ &$0$ & 0&0\\
$0$ & $\mathbf{ 0}$ &$0$ &$.426$ & .811&.689&.426 & $0$ & $0$ &$0$ &$0$ & 0&0\\
$0$ & $0$ &$.426$ &$.891$ & .703&.583&.426 & $0$ & $0$ &$0$ &$0$ &0&0\\
$0$ & $.426$ &$\mathbf{.937}$ &$\mathbf{.700}$ &$\mathbf{ .510}$&.586&.426 & $0$ & $0$ &$0$ &$0$ & 0&0\\
$.426$ & $.964$ &$.701$ &$\mathbf{.518}$ & .588&.635&.426 & $0$ & $0$ &$0$ &$0$ & 0&0\\
\ldots
\end{tabular}
\end{center}
\caption{Evolution of a nonlinear rule with initial seed $a=0.426$ (truncations of actual values used)}
\label{t4}
\end{table}

In this sense the initial string now consists of a set of fuzzy states, that is a collection of arbitrary but fixed real numbers in the closed interval $[0, 1]$. Inherent in this procedure is the fact that fuzzification allows one to move from the discrete (boolean) to the continuous (fuzzy) by extending the domain of definition of the rule (by modifying the underlying multi-valued logic). We now turn to the process of {\it fuzzifying} boolean rule 110 under probabilistic logic.

\begin{Exa}\label{proba} (Probabilistic rule 110 or PCA 110) For real numbers $a,b \in [0,1]$, in the DNF for rule 110 we define the quantities $(a \vee b)$   to be $(a+b-a\cdot b)$, $(a \wedge b)$ to be $(ab)$, and $(\neg a)$ to be  $(1-a)$, in accordance with the probabilistic logic above. In other words, $a \vee b = a+b - a\cdot b$, $a \wedge b = a\cdot b$, and $\neg a = 1-a$, where $+$ and ``$\cdot$" are ordinary addition and multiplication of real numbers. As before, using the DNF of ECA 110 as a starting point, we find 
\begin{eqnarray}
g^{P}_{110}(x_1,x_2,x_3) & = & (\neg x_1 \wedge  \neg x_2 \wedge x_3) \vee (  \neg x_1 \wedge  x_2 \wedge  \neg x_3) 
\vee ( \neg x_1 \wedge x_2 \wedge  x_3) \vee \nonumber \\
&& \vee (  x_1 \wedge \neg x_2 \wedge  x_3) \vee (  x_1 \wedge x_2 \wedge  \neg x_3) \nonumber \\
& = & (1-x_1)(1-x_2)x_3(1-A) + A\nonumber 
\end{eqnarray}
\end{Exa}
where $A=(1-x_1)x_2(1-x_3)(1-B) + B,$ $B=(1-x_1)x_2x_3(1-C) + C,$ and $C=x(1-y)z + xy(1-z)-x^2yz(1-y)(1-z).$  Written out in full this becomes:


\begin{eqnarray*}
\lefteqn{g^{P}_{110}(x_1,x_2,x_3) = }\\
&& (1-x_1)(1-x_2)x_3(1-(1-x_1)x_2(1-x_3)(1-(1-x_1)x_2x_3 (1- \\
&& x_1(1-x_2)x_3 - x_1x_2(1-x_3)+x_1^2x_2x_3(1-x_2)(1-x_3))- \\ 
&& x_1(1-x_2)x_3- x_1x_2(1-x_3) + x_1^2x_2x_3(1-x_2)(1-x_3))- \\ 
&& (1-x_1)x_2x_3(1-x_1(1-x_2)x_3-x_1x_2(1-x_3) + x_1^2x_2x_3(1- \\ 
&& x_2)(1-x_3))-x_1(1-x_2)x_3-x_1x_2(1-x_3)+x_1^2x_2x_3(1-x_2)(1-x_3)) \\
&& + (1-x_1)x_2(1-x_3)(1-(1-x_1)x_2x_3(1-x_1(1-x_2)x_3-x_1x_2(1-x_3)\\ 
&& + x_1^2x_2x_3(1-x_2)(1-x_3))-x_1(1-x_2)x_3- x_1x_2(1-x_3) \\ 
&& +x_1^2x_2x_3(1-x_2)(1-x_3)) + (1-x_1)x_2x_3(1-x_1(1-x_2)x_3- \\ 
&& x_1x_2(1-x_3)+ x_1^2x_2x_3(1-x_2)(1-x_3)) + x_1(1-x_2)x_3\\ 
&& + x_1x_2(1-x_3)-x_1^2x_2x_3(1-x_2)(1-x_3), \\ \\
\end{eqnarray*}
or, equivalently, 
\begin{eqnarray*}
&&\lefteqn{g^{P}_{110}(x_1,x_2,x_3) = }\\
&& x_2+x_3-x_1x_2x_3-3\,x_1^2x_2x_3^2-3\,x_1^2x_2^2x_3+5\,x_1^2x_2^2x_3^2+2\,x_1^3x_2^2x_3^2 \\ 
&& +4\,x_2x_3^2x_1-9\,x_2^2x_3^2x_1+ 4\,x_2^2x_3x_1+5\,x_2^2x_3^3x_1^2+5\,x_2^3x_3^2x_1^2- 13\,x_2^3x_3^3x_1^2- \\
&& 7\,x_2^2x_3^3x_1^3- 7\,x_2^3x_3^2x_1^3+ 11\,x_2^3x_3^3x_1^3- 2\,x_2x_3-x_1x_3^2+x_1^2x_3^2-x_1x_2^2\\ 
&& +x_1^2x_2^2 +2\,x_2^2x_3^2- x_2^3x_3^2-x_2^2x_3^3+x_2^3x_3^3+x_2^2x_3^3x_1- 3\,x_2^2x_3^4x_1^2\\ 
&& +6\,x_2^3x_3^4x_1^2-x_1^2x_2x_3^3+3\,x_2^2x_3^4x_1^3-x_2^3x_3^4x_1^3 -x_1^4x_2^2x_3^2+ 2\,x_2^2x_3^3x_1^4 \\
&& +2\,x_2^3x_3^2x_1^4- x_2^3x_3^3x_1^4+x_1^3x_3^3x_2-x_1^4x_3^4x_2^2-4\,x_1^4x_3^4x_2^3-x_1^2x_2^3x_3 \\ 
&& +x_1^3x_2^3x_3+x_2^3x_3^2x_1- 3\,x_2^4x_3^2x_1^2+ 6\,x_2^4x_3^3x_1^2+3\,x_2^4x_3^2x_1^3-x_2^4x_3^3x_1^3\\ 
&&+3\,x_2^3x_3^3x_1-x_2^4x_3^4x_1^2-8\,x_2^4x_3^4x_1^3-x_2^4x_3^2x_1^4-4\,x_2^4x_3^3x_1^4+11\,x_1^4x_3^4x_2^4\\ 
&& +x_2^4x_3^2x_1+x_2^5x_3^3x_1^2-3\,x_2^5x_3^3x_1^3-3\,x_2^4x_3^3x_1 - 2\,x_2^5x_3^4x_1^2+6\,x_2^5x_3^4x_1^3\\ 
&& +3\,x_2^5x_3^3x_1^4-6\,x_1^4x_3^4x_2^5-x_2^3x_3^3x_1^5+2\,x_1^5x_3^4x_2^3+2\,x_2^4x_3^3x_1^5- \\ 
&& 4\,x_1^5x_3^4x_2^4- x_1^5x_2^5x_3^3+2\,x_1^5x_2^5x_3^4+x_2^2x_3^4x_1+x_2^3x_3^5x_1^2-3\,x_2^3x_3^5x_1^3\\ 
&& +3\,x_1^4x_3^5x_2^3- 3\,x_2^3x_3^4x_1-2\,x_2^4x_3^5x_1^2+6\,x_2^4x_3^5x_1^3-6\,x_1^4x_3^5x_2^4+2\,x_2^4x_3^4x_1\\ 
&& +x_2^5x_3^5x_1^2- 3\,x_2^5x_3^5x_1^3 +3\,x_1^4x_3^5x_2^5-x_1^5x_3^5x_2^3+2\,x_1^5x_3^5x_2^4-x_1^5x_2^5x_3^5.
\end{eqnarray*}
Now the {\it look up table} of a rule $g$ on $[0,1]^3$ is the image of the 8 boolean triples $(0,0,0), (0,0,1), \ldots, (1,1,1)$ under $g$. Given the nature of probabilistic ECA 110 (PCA 110) as seen above, we observe that the look up tables of probabilistic and FCA 110 are identical. In other words, the values of the corresponding fuzzification $g(x,y,z)$ all agree on the 8 basic triples. Although more difficult to write down explicitly, it can be verified directly without too much effort that this is also the case for Product ECA 110 (see Table~\ref{ta1} below), G\"{o}del ECA 110 and even the Lukasiewicz ECA 110 (for any $p \geq 1$). In addition, these continuous ECA also agree with Boolean ECA 110 when restricted to the Boolean space $\{0,1\}$. In fact, the reason for this equivalence of look-up tables, for any rule number, is that all the multi-valued logics mentioned here agree with standard propositional calculus when restricted to the two-point set $\{0,1\}$.

\begin{table}
\begin{center}
\begin{tabular}{||c|c||c|c|c||}\hline
Boolean triples\, &\, Boolean 110\, &\,  Fuzzy 110\, &\,  Probabilistic 110\,&\, Product 110\, \\ \hline
$000$&$	0	 $&$ 0			$&$ 0			$&$ 0			$ \\
$001$&$	1	 $&$ 1			$&$ 1			$&$ 1			$ \\
$010$&$	1	 $&$ 1			$&$ 1			$&$ 1			$ \\
$011$&$	1	 $&$ 1			$&$ 1			$&$ 1			$ \\
$100$&$	0	 $&$ 0			$&$ 0			$&$ 0			$ \\
$101$&$	1	 $&$ 1			$&$ 1			$&$ 1			$ \\
$110$&$	1	 $&$ 1			$&$ 1			$&$ 1			$ \\
$111$&$	0	 $&$ 0			$&$ 0			$&$ 0			$ \\
\hline
\end{tabular}
\end{center}
\caption{The look-up tables of the elementary cellular automaton, ECA 110, under diverse multi-valued logics}
\label{ta1}
\end{table}

\section{The long term dynamics of ECA under CFMS logic}
\begin{Exa}
\label{ex184}
Since $184 = 2^3+2^4+2^5+2^7$ we see that the rule number $184 = \sum_{i=0}^{7} r_i\,2^i $
forces $r_i = 1$ only for $i=3, 4, 5, 7$. Use of the disjunctive normal form expression above gives us
\begin{eqnarray}
g_{184}(x_1,x_2, x_3) & = &\vee_{i|r_i=1} \wedge_{j=1}^{3} x_j^{d_{ij}},\nonumber\\
&=&  (x_1^{d_{31}} \wedge x_2^{d_{32}} \wedge x_3^{d_{33}}) \vee (x_1^{d_{41}} \wedge x_2^{d_{42}} \wedge x_3^{d_{43}}) \vee  (x_1^{d_{51}} \wedge x_2^{d_{52}} \wedge x_3^{d_{53}})\nonumber \\
&& \vee (x_1^{d_{71}} \wedge x_2^{d_{72}} \wedge x_j^{d_{73}}), \nonumber \\
& = & (x_1^{{0}} \wedge x_2^{{1}} \wedge x_3^{{1}}) \vee (x_1^{{1}} \wedge x_2^{{0}} \wedge x_3^{{0}}) \vee (x_1^{{1}} \wedge x_2^{{0}} \wedge x_3^{{1}}) \vee (x_1^{{1}} \wedge x_2^{{1}} \wedge x_3^{{1}}), \nonumber \\
& = & (\neg x_1 \wedge  x_2 \wedge x_3) \vee (  x_1 \wedge  \neg x_2 \wedge  \neg x_3) 
\vee ( x_1 \wedge \neg x_2 \wedge  x_3) \vee \nonumber \\
 && \vee (  x_1 \wedge  x_2 \wedge  x_3) \label{b11}\\
&=& (1-x_1)x_2x_3 + x_1(1-x_2)(1-x_3) + x_1(1-x_2)x_3 + x_1x_2x_3,\label{f10}\\
& = & x_1-x_1x_2+x_2x_3.\label{f1bis}
\end{eqnarray}
\end{Exa}
Note that in ``fuzzifying" the DNF (\ref{b1}),  we replaced $\neg x$ by $1-x$, $ x \vee y$ by $x+y$, and $x \wedge y$ in (\ref{b1}) by their product, $xy$, so as to find the fuzzy form (or rule) given by (\ref{f10}) or equivalently (\ref{f1bis}), in accordance with the fuzzification process defined above. In this case, the local rule maps the triples of zeros and ones as follows:
$$000, 001, 010, 011, 100, 101, 110, 111 \rightarrow
0,0,0,1,1,1,0,1.$$ Thus, (boolean) rule 184 is given by (\ref{b11}) above while FCA 184, given by (\ref{f1bis}), may be written as
$$g_{184}(x,y,z) = x - xy + yz,$$ for any value of $(x,y,z) \in [0,1]^3,$  (the unit cube in $\mathbf{R}^3$.) From now on, ``rule $n$" refers simply to the local ECA correponding to (classical) boolean cellular automata defined above (or in \cite{nks}) while ``FCA $n$" is the CFMS-fuzzification of the local rule as described above. We emphasize that to each classical boolean CA corresponds a {\it unique} CFMS CA and {\it vice versa}.

\begin{Exa}
\label{ru18}
Rule $18 = 2 + 2^4$ has the local rule
$$(000, 001, 010, 011, 100, 101, 110, 111) \rightarrow
(0,1,0,0,1,0,0,0).$$
Its canonical expression being
$$
g_{18}(x_1,x_2, x_3) =(\neg x_1 \wedge \neg x_2 \wedge
x_3)
\vee ( x_1 \wedge \neg x_2 \wedge \neg x_3),
$$
we obtain its fuzzification as%
\begin{eqnarray}
\label{r18}
g_{18}(x_1,x_2,x_3) = (1-x_2) \cdot (x_1 + x_3 - 2x_1
\cdot x_3).
\end{eqnarray}
\end{Exa}

One of the ECA that motivated this paper follows.

\begin{Exa}
\label{ex30}
Rule $30 = 2 + 2^2+2^3+2^4$ has the local rule
$$(000, 001, 010, 011, 100, 101, 110, 111) \rightarrow
(0,1,1,1,1,0,0,0).$$
Its canonical expression is
$$
g_{30}(x_1,x_2, x_3) =(\neg x_1 \wedge \neg x_2 \wedge
x_3)
\vee ( \neg x_1 \wedge  x_2 \wedge \neg x_3)
\vee ( \neg x_1 \wedge  x_2 \wedge  x_3) \vee ( x_1
\wedge  \neg x_2 \wedge  \neg x_3)
$$
so its fuzzification becomes
\begin{eqnarray}
\label{r30}
g_{30}(x_1,x_2,x_3) = x_1+x_2+x_3
-2x_1x_2-x_2x_3-2x_1x_3+2x_1x_2x_3.
\end{eqnarray}
\end{Exa}

The usual fuzzification of the expression $a \vee b$
is
$max\{1, a+b\}$ so as to ensure that the result is not
larger than $1$. 
Note, however, that taking $(a+b)$ for the CA
fuzzification does not
lead to values greater than $1$ since the sum of all 
the expressions for rule 255 is $1$ ({\em i.e.,}
$g_{255}(x,y,z)=1$), 
and so every (necessarily non-negative) partial sum
must be bounded by $1$. Since every FCA is
obtained by adding one or more of these partial sums
it follows that every FCA is bounded below by 0
and above by 1.

Indeed, we note that there are many ways of fuzzifying a rule so that this process of fuzzification is not unique by any means (see \cite{reiter} for other possible fuzzifications). In particular, we mention one type of unusual fuzzification where the value $a \vee b$ is replaced by $1- (1-a)(1-b)$. If adopted, this would lead to another continuous rule where considerations similar to those in \cite{abmf}, \cite{ijuc} may be applied to deduce rule asymptotics. In order to distinguish this definition from other possible definitions in the literature (e.g., \cite{reiter}) we sometimes add the prefix CFMS (an acronym) in honor of its discoverers, \cite{cat1}. In order to make this report as self-contained as possible we will define an {\it exceptional fuzzy cellular automaton} herewith: 

\begin{Defi} \label{def2} A (CFMS CA) or FCA, for short, with local rule $g$ is said to be {\rm exceptional} provided its diagonal function $d$, is identically zero on $[0,1]$ (where $d(x) \equiv g(x,x,x) -x$). That is, the set of fixed points of $g$ is the closed interval $[0,1]$.
\end{Defi}

A complete analysis of all 255 (CFMS) FCA's shows that there only nine such exceptional fuzzy CA, that is, FCA 170, 172, 184, 202, 204, 216, 226, 228, and 240.   The definition of 
these nine exceptional FCA, of which six are non-trivial, indicate
that the general methods presented in \cite{abmf}, \cite{ijuc} require major
modifications. Of these nine, three are trivial, that is fuzzy ECA 170, 204 and 240 have simple dynamics, for any initial string, finite or infinite. The point is that all but possibly these six remaining FCA have non-chaotic dynamics at infinity (compare \cite{cat2}). These inherent difficulties in the remaining six FCA were treated recently in a series of separate papers \cite{iccs} and \cite{acri}, thus completing the program of the classification of the dynamics of all 255 FCA under fuzzy logic.

To begin with, we fix the notation. We recall some
definitions from \cite{rule90}. The {\it space-time diagram}
from a cell $x_{i}^{t}$ is the set of $\{ x_j^{t+p}
\mid
  p\geq 0\, {\rm{and\,}} j \in \{ i-p, \cdots , i+p \}
\}$. In the case of a single fuzzy seed, $a$, in a background of zeros (i.e., its neighbors are all zero), the space-time diagram is the infinite triangle whose vertex is at the singleton
$a=x_{0}^{0}$. Thus, $x_{\pm n}^{m}$ will denote the cell at
$\pm n$ steps to the right/left of $a$ (the zero state) at
time $m$. For example, $x_{-1}^3 =0.5429, x_{1}^{2}=0,
x_{-2}^2 = 0.875, \ldots$ in the adjoining 
Table~\ref{rule110} for the evolution of FCA 110 starting from a single seed $a=0.5$ in a background of zeros. The single cell $x_{0}^{0}$
will always be denoted by $a$ and generally we will take it
that $0 < a \leq 1$, since $a=0$ leads to trivial asymptotics.

\begin{table}[ht]
  \begin{displaymath}
    \begin{array}{c|ccccccccc}
      & \multicolumn{9}{c}{\mbox{Local states}}\\
      \mbox{Time} & \cdots & -3 & -2 & -1 & 0 & 1 & 2
& 3 & \cdots\\
      \hline \\
\cdots & \cdots \\ \\
      1 & \cdots & 0 & 0 & .5 & .5 & 0 & 0 & 0 &
\cdots\\ \\
      2 & \cdots & 0 & .5 & .75 & .5 & 0 & 0 & 0 &
\cdots\\ \\
      3 & \cdots & .5 &.875& .6875 &.5 & 0 & 0& 0 
      & \cdots\\ \\
      4  & \cdots &.9375 & .6601 & .5429& .5 & 0&0& 0 &
\cdots\\ \\
    \end{array}
  \end{displaymath}
  \caption{\label{rule110}
   FCA 110: Partial evolution from the point $a=0.5$ in a zero background.}
\end{table}

By way of further motivation for the study of the global evolution or temporal asymptotics of fuzzy ECA we assume that the initial string is the seed $a=1$, in a background of zeros. For a given integer $n$ where $1 \leq n \leq 255$ it is readily verified that the space-time diagrams of both the fuzzy and boolean rule n are {\it identical}. In other words, restricting our given FCA $n$ to the case where $a=1$, the resulting evolution of $0$'s and $1$'s in its space-time diagram will necessarily produce an output that is identical to that which we would obtain had we applied (boolean) rule $n$ to our initial string. The reason for this is that (the name of) a FCA is uniquely determined by its action on the eight elements $000, 001, \ldots$ so one need only calculate these eight values using the fuzzy local function and compare them to the corresponding boolean values obtained by using the DNF.

In addition, proceeding from a single seed in a background of zeros again, the space-time diagram produced by applying FCA $n$ to this string will ``converge" to the boolean one as $a \to 1^-$. In this way we show that complexifying the problem by introducing a fuzzy seed (as opposed to a boolean one) can actually lead to a somewhat deeper understanding of the evolution of a given boolean rule. 

In this respect we use the following device (see \cite{rule90}).  Fix a FCA and choose an initial seed $a=x_{0}^{0}$, in a zero background such that $a$ is very close to, but less than, the number $1$. We then fix a row number or, equivalently, a specified time $T >0$.  Consider the fuzzy evolution of the rule up to time $T$. Thus, we are looking at the light-cone proceeding from $a$ to the row whose elements are $x_{-T}^{T}, x_{-T+1}^{T}, \ldots, x_{0}^{T}, \ldots, x_{T+1}^{T}, x_{T}^{T}.$ Because of the results in \cite{abmf}, \cite{ijuc} on the determinacy of the asymptotic limits of all but six FCA we have that the limits are continuous functions of the seed $a$. In addition, we recall that both boolean and fuzzy space-time diagrams are identical if $a=1$. We now refer to the fuzzy space-time diagram:  If any cell at time no larger than $T$ has an entry {\it greater} than $1/2$ we color it {\it black}. On the other hand, if a cell at time no larger than $T$ has an entry {\it less} than $1/2$ we color it {\it white}.  The result is that if the seed $a$ is chosen sufficiently close to $1$, then the space-time diagrams of both the boolean and the fuzzy ECA up to time $T$ will be {\it identical}! Thus, one can expect that the asymptotic evolution of the fuzzy space-time diagram will have bearing on the distribution of $0$'s and $1$'s in the corresponding asymptotic  boolean evolution. 

This method is illustrated in Table~\ref{t1} and Table~\ref{t2} below and adapted to the space-time diagram for FCA 110, emanating from a single seed, for simplicity. We recall that the local function for this rule is given in \cite{amca1}, or that $g_{110}(x,y,z) = y+ z - yz -xyz$, for any $x,y,z \in [0,1]$. Thus, $g_{110}(.99,.09,.95) = .86$, $g_{110}(.14, .20, .25) = .4$, etc.

\begin{table}{}

\begin{center}


\begin{tabular}{ c  c  c  c c  c  c c}	\hline

$0$ & $0$ &$0$ &$0$ & 0&.95&.95 & 0 \\	\hline

$0$ & $0$ &$0$ &$0$ & .95&.99&.95 & 0  \\

$0$ & $0$ &$0$ &$.95$ & .99&.09&.95 & 0\\

$0$ & $0$ &$.95$ &$.99$ & .90&.86&.95 & 0\\

$0$ & $.95$ &$.99$ &$.14$ & .20&.25&.95 & 0\\

$.95$ & $.99$ &$.86$ &$.28$ & .40&.91&.95 & 0\\

$.99$ & $.17$ &$.65$ &$.47$ & .84&.64&.95& 0 \\

\ldots

\end{tabular}


\end{center}

\caption{Partial space-time diagram of FCA 110 from $a=0.95$ up to $T \leq 7$.}

\label{t1}

\end{table}

\begin{table}{}

\begin{center}


\begin{tabular}{rrrrrrr}	\hline

$0$ & $0$ &$0$ &$0$ & 0&0&1\\	\hline

$0$ & $0$ &$0$ &$0$ & 0&1&1   \\

$0$ & $0$ &$0$ &$0$ & 1&0&1 \\

$0$ & $0$ &$0$ &$1$ & 1&1&1 \\

$0$ & $0$ &$1$ &$0$ & 0&0&1 \\

$0$ & $1$ &$1$ &$0$ &0&1&1 \\

$1$ & $0$ &$1$ &$0$ & 1&1&1\\

\hline

\end{tabular}

\end{center}

\caption{Partial space-time diagram of boolean ECA 110 from $a=1$ up to $T \leq 7$.}

\label{t2}

\end{table}

\begin{remark} Note that we truncate the values of the cells in Table~\ref{t1} above to two decimal places since we are only interested in values relative to the number $1/2$. Once that cell value is greater than $1/2$ we color it black (or give it the value $1$) and if it is less than $1/2$ we color it white (or give it the value $0$). In this way we see that the space-time diagrams described in Tables~\ref{t1}-\ref{t2} are identical.
\end{remark}

\begin{remark} At this time we cannot decide the color of a fuzzy cell whose value is exactly one-half. Although this cell value equal to $1/2$ is unlikely in the case of random inputs of the initial seed $a$, it can occur if, say, the seed value $a=1/2$ (such as in FCA 90, \cite{rule90}). Indeed, it is a consequence of the definition of fuzziness here that given any one of the 256 fuzzy ECA, the closer the seed value $a$ is to $1$ the more ``agreement" there is between the boolean and fuzzy space-time diagrams, using the device mentioned above that allows the transition from a fuzzy scenario to a boolean one. Consequences of this remark will be addressed at a later time in a different work. 
\end{remark}

\subsection{The Dynamics of FCA 110}
\label{0110}
Interest in the boolean form of FCA 110 has
been tremendous (see \cite{nks}) since Matthew Cooke's
announcement that it can support Turing-complete
computation. We use the general theory developed in \cite{abmf}, \cite{ijuc} to derive the dynamics
of the fuzzification of this rule and so provide a
simple proof of the results obtained in \cite{amca1}. We note
that the long term dynamics of FCA 110 were
obtained previously in \cite{amca1} using special arguments
pertaining to the form of the rule itself along with
its representation as a Taylor polynomial, reminiscent
of the techniques used earlier in \cite{rule90}.

The canonical expression for this rule is given by
$$g_{110} (x,y,z) = y+z-yz-xyz.$$
The fixed points in $\mathcal{U}$ of this FCA
are $x=0$, repelling, $x=(\sqrt{5}+1)/2$, which lies
outside the domain of the rule and finally, the golden
number, $x =(\sqrt{5}-1)/2$ which is attracting. The
dynamics of this rule are given by solving the equations
\begin{eqnarray*}
L_{m+1}^{-}(a) &=& g_{110} \left (L_{m-1}^{-}(a),
L_{m}^{-}(a), L_{m+1}^{-}(a)\right) 
\end{eqnarray*}
using the initial conditions
$L_{0}^{-}=a, L_{1}^{-}=1,$ (see \cite{ijuc} for a proof). So, all  limits are found
recursively with the first few 
limits being given by
$$L_{2}^{-} = \frac{1}{1+a},\,\,
L_{3}^{-}=\frac{1}{2}, \,\, L_{4}^{-} =
\frac{a+1}{a+2}, \,\,L_{5}^{-} = \frac{2}{3},
\,\,L_{6}^{-} = \frac{a+2}{2a+3}, \,\,L_{7}^{-}=
\frac{3}{5},$$ and $$L_{8}^{-} = \frac{2a+3}{3a+5}, \,\,
\ldots . $$
An induction argument shows that the even and odd
subscripted left-diagonal limits $L_{2n}^{-}$ and
$L_{2n-1}^{-}$ are given more simply by 
\begin{eqnarray}
\label{fibo}
L_{2n}^{-}& = & \frac{1}{1+L_{2n-2}^{-}},\,\,\,\,\,\,
L_{2n+1}^{-} = \frac{1}{1+ L_{2n-1}^{-}},
\end{eqnarray}
for each $n \geq 1$. 

\begin{Rem}
As an aside we note that the second formula in
(\ref{fibo}) defines the continued fraction expansion
(cf., \cite{wall}) of the golden number, $\phi \equiv
(\sqrt{5} -1)/2 = 0.618033\ldots$. As a result, the
fractional representation of  $L_{2n+1}^{-}$ appearing
in (\ref{fibo}) must be quotients of Fibonacci numbers
(viz., $L_{2n-1}^{-} = {F_n}/{F_{n+1}}$). where $F_n$
is the $n^{th}$ Fibonacci number, $F_0=0$. It follows
that the golden number is a limit point of the
infinite sequence of left-diagonal limits in the
evolution of FCA 110. Next, one can show by
induction that 
\begin{eqnarray}
\label{fibo2}
L_{2n}^{-}& = & \frac{a\,F_{n-1}+F_n}{a\, F_n +
F_{n+1}},
\end{eqnarray}
for each $n \geq 1$. 
\end{Rem}
Factoring out a copy of $F_n$ from both numerator and
denominator in (\ref{fibo2}), passing to the limit as
$n \to \infty$ and using standard limiting properties
of quotients of Fibonacci numbers we see that
$L_{2n}^{-} \to \phi$, independently of the seed, $a$.
We can therefore conclude that the limit of the
left-diagonal limits in FCA 110 actually exists
and is equal to the golden number (see \cite{amca1} for more details).
\begin{Rem} These results about the limits of FCA 110 indicate that there is some ``order" to the evolution of the space-time diagram of this specific CA. However, since each such limit is different from the preceding one this also implies that, generally speaking, the limit of a sequence of the form $\{x_{n}^{m}\}$ in its space-time diagram will NOT exist at infinity. There are FCA for which even these arbitrary limits DO exist, but they are few and shall be taken up in a future paper.
\end{Rem}

\section{The dynamics of PCA 110}

The most difficult investigation using any means whatsoever appears in this section. Here we begin with ECA 110 and change the underlying logic to a probabilistic one (see Example~\ref{proba} for its local rule) and proceed to study the dynamics of its space-time diagram. First we note that the diagonal function of ECA 110 under probabilistic logic is given by the polynomial (see Example~\ref{proba} above) of degree 15,
\begin{eqnarray*}
g_{110}^{P}(x,x,x) &=& -2\,{x}^{2}-21\,{x}^{13}+7\,{x}^{14}-{x}^{15}-40\,{x}^{8}-27\,{x}^{11} +34\,{x}^{12}\\
&& +6\,{x}^{6}+35\,{x}^{9}-4\,{x}^{10}+19\,{x}^{7}+2\,x-3\,{
x}^{3}+12\,{x}^{4}-17\,{x}^{5}.
\end{eqnarray*}
Its fixed points correspond to the zeros of $g_{110}^P(x,x,x)-x$, that is, the solutions of 
\begin{eqnarray*}
x-2\,{x}^{2}-3\,{x}^{3}+12\,{x}^{4}-17\,{x}^{5}+6\,{x}^{6}+19\,{x}^{7}
-40\,{x}^{8}+35\,{x}^{9}-4\,{x}^{10}- \\ 
27\,{x}^{11}+34\,{x}^{12}-21\,{x}
^{13}+7\,{x}^{14}-{x}^{15}=0.
\end{eqnarray*}
By inspection we note that $x=0$ is a root and there is an irreducible factor $x^2-x+1$ in the display above. It follows using the theory of {\it Sturm sequences} [\cite{burn}, p.198] that this diagonal function has exactly three real roots only two of which are within $[0,1]$ (including the one at $x=0$) the other being at approximately $x_0=0.4845$. It is readily verified that there is a repelling fixed point at $x=0$ and an attracting fixed point at $x=x_0$ (the precise value of which is unknown).

We now apply PCA 110 to the case of a single (fuzzy) seed $a \in (0,1)$ in a background of zeros (see Section~\ref{0110} above in the case of FCA 110). The nature of this cellular automaton based on probabilistic logic makes it such that $g_{110}^{P}(x,y,z) $ introduced earlier is now an algebraic function (specifically, a polynomial in three variables of degree 5 in each variable separately). Still, it is possible to a certain extent to determine the long term behavior of the diagonals $L_m^{\pm}$ for this specific rule. For example, let us proceed as in \cite{ijuc} in the single seed case, where we choose and fix a seed $a \in (0,1)$. Observe that the space-time diagram of PCA 110 is similar in {\it shape} to the FCA 110 case (no evolution to the right of the central column with apex at $x_0^0$). Since $g_{110}^{P} : [0,1]^3 \to [0,1]$ is continuous on this domain, the method in \cite{abmf} applies to give the limits recursively. Thus, noting that $L_0^{-}=0$ and $L_{1}^{-}=a$, we find that
\begin{eqnarray*}
L_{2}^{-}(a) &=& g \left (L_{0}^{-}(a),
L_{1}^{-}(a), L_{2}^{-}(a)\right) \\
& = &  g \left (0, a, L_{2}^{-}(a)\right) \\
& = & a+{ L_{2}^{-}(a)}-2\,a{ L_{2}^{-}(a)}+2\,{a}^{2}{({ L_{2}^{-}(a)})}^{2}-{a}^{3}{({ L_{2}^{-}(a)})}^{2} -{a}^{2}{({ L_{2}^{-}(a)})}^{3} \\ 
&& +{a}^{3}{({ L_{2}^{-}(a)})}^{3},
\end{eqnarray*}
a cubic with real coefficients in the unknown $L_{2}^{-}(a)$. For a non-trivial seed value (i.e., $a \neq 0$) our limit  $L \equiv L_{2}^{-}(a)$ must satisfy the cubic equation
\begin{eqnarray}\label{big} p(L) \equiv (a^2-a)L^3 +(2a-a^2)L^2 - 2L +1 =0.
\end{eqnarray}
However, since $p(0)p(1)<0$ and $a \in (0,1)$, Bolzano's theorem guarantees for each $a$ the existence of a real root $x_0(a) \equiv L(a) \in (0,1)$. For such an $a \in (0,1)$, an application of the implicit function theorem indicates that $L_2^{-}$ is a continuous function of $a$ at least locally. Varying the seed we get that  $L_2^{-}$ is continuous for every $a \in (0,1)$. 

Chaotic evolution may occur if other zeros of $p(L)$ present themselves in $(0,1)$, since then it may be conceivable that seeds close to one another may converge to different limits under the action of PCA 110. However, we show presently that this is not the case, that is, for any given value of the seed $a \in (0,1)$ the cubic polynomial $p(L)$ above has exactly one root in $(0,1)$. To this end let $x_0(a)$ denote a real root of $p(L)$. We follow the root as $a$ varies. Since
\begin{eqnarray}\label{ax0}
(a^2-a){x_0^3(a)} +(2a-a^2)x_0^2(a) - 2x_0(a) +1 =0,
\end{eqnarray}
the implicit function theorem gives that $x_0$ is a differentiable function of $a$ and, differentiating both sides with respect to $a$, we obtain after some simplification,
\begin{equation}\label{x0a}
\bigg (3ax_0^2(a)(a-1)+2ax_0(a)(2-a) - 2\bigg )\frac{dx_0}{da} + x_0^2(a)\left ( (2a-1)x_0(a) +2-2a\right ) = 0.
\end{equation}
First note that the root cannot be stationary and, in fact, the motion of this root is monotone, for otherwise, we must have $dx_0/da = 0$ for some $a$ which, along with (\ref{x0a}) forces $x_0 = (2a-2)/(2a-1)$. However, since this $a \in (0,1)$, this last equation is not possible for our root in $(0,1)$ and this establishes our result.

Now we prove the uniqueness of the real root $x_0(a) \in (0,1)$. If possible, let $p(L)$ have two real roots; then clearly, all its roots $x_0, x_1, x_2$ must be real. The classical relationship between these  roots and the coefficients of (\ref{ax0}) gives the result
\begin{eqnarray}
\label{x0x1}
2x_0x_1x_2=x_0x_1+x_1x_2+x_0x_2.
\end{eqnarray}
An application of the arithmetic-geometric inequality $(A+B+C)/3 \geq (ABC)^{1/3},$ (see \cite{hlp}) for the case of three positive numbers $A=x_0x_1$, $B=x_1x_2$, $C=x_0x_2$ along with (\ref{x0x1}) gives the impossible result $8x_0x_1x_2\geq 27$ since all $x_i <1$. It follows that no two real roots can exist for (\ref{big}) and so the real root $x_0(a)$ found above is unique. The consequence of all this argumentation is that the limit $L_2^-(a)$ is well-defined, continuous and unique (it can be found numerically using Newton iterations) and thus no sensitive dependence (see \cite{yor}) on the initial seed $a$ can occur for this diagonal. It is likely the case that the same holds true for the limits of all subsequent diagonals but we cannot prove this (due to the huge number of terms that present themselves). Nevertheless, we suspect that chaos cannot occur in the evolution of a space-time diagram from a single seed in a homogeneous background of zeros for PCA 110 although we note that this apparent lack of complexity seems to have no bearing on whether PCA 110 is universal or not (the answer to which is unknown to us).

\section{Concluding Remarks}

Now that the program of investigating the dynamics of all fuzzy cellular automata initiated in \cite{abmf}, \cite{ijuc} is completed with the appearance of  \cite{iccs}, \cite{acri}, attention is turned to studying the dynamics of these 255 ECA relative to modification of the underlying logical structures. The so-called boolean case (induced by a two-valued logic) has been studied extensively beginning with the works of Wolfram (see e.g., \cite{nks} and the references therein). Specializing to ECA 110 we show that although boolean ECA 110 is universal, its fuzzy counterpart, FCA 110, cannot be as its space-time diagram cannot be chaotic. In addition, modifying the underlying logic to a probabilistic one, as we have done in this paper, the techniques introduced  indicate that such asymptotic data can be calculated at least implicitly if not explicitly, the surprising result being that even a probabilistic logic applied to ECA 110 seems to exhibit non-chaotic behavior! To what extent chaos is a part of boolean CA's under ordinary logic is uncertain, but it appears that some CA's under continuous logics may surprisingly exhibit non-chaotic space-time diagrams.
\vskip0.15in
\noindent{{\bf Acknowledgments:}} This research is partially supported by an NSERC
Canada Research Grant and by a grant from the Office
of the Vice-President Research and International,
Carleton University. Gratitude is also expressed to the Department of Mathematics, University of Las Palmas, Gran Canaria, Spain, for their hospitality and where part of this work was completed. The author also thanks Professor N. Santoro (Carleton University) and Professor P. Flocchini (University of Ottawa) for a helpful discussion.

\end{document}